# A Methodology for Implementation of MMS Client on Embedded Platforms

A. A. Milani[1], M. R. Rahimi[2]

**Abstract**

MMS (**M**ultimedia **M**essaging **S**ervice) is the next generation of messaging services in multimedia mobile communications. MMS enables messaging with full multimedia content including images, audios, videos, texts and data, from client to client or e-mail. MMS is based on WAP technology, so it is technology independent. This means that enabling messages from a GSM/GPRS network to be sent to a TDMA or WCDMA network. In this paper a methodology for implementing MMS client on embedded platforms especially on Wince OS is described.

## 1. Introduction

In this article the overall architecture for MMS SMIL player and composer according to the MMS conformance document version 2.0.0 is described. SMS (**S**hort **M**essage **S**ervice) is the first messaging system in mobile communication that could just send simple static text or image, but MMS not only could contain several media types (text, image, audio and video) but also by using a presentation language could put these media on a sequential basis and play the message. The playback consist of several slides (each slide contains its own text, image, audio and video) which are presented according to a time table that shows the start and the duration time of each slide relative to the message playback start time. This presentation language is called SMIL.

SMIL (**S**ynchronized **M**ultimedia **I**ntegration **L**anguage, which is pronounced smile) is the Markup presentation language, accepted and standardized by industry for media applications. SMIL's focus isn't on data encoding, it does not specify CODECs (decoding engine and formats) for any media type such as video or audio but on focuses on media integration. SMIL specifies how components are related temporally and spatially during a presentation [1, 2, 3].

Here a MMS SMIL player and composer architecture which is independent of the media formats is described. To be a platform independent solution and the independency between media CODECs and this engine, makes this architecture a suitable solution for MMS systems.

---

[1] Ali_a_milani@yahoo.com, Eimaa Communication Technologies.
[2] rahimi_mr@yahoo.com, Sharif University of Technology.

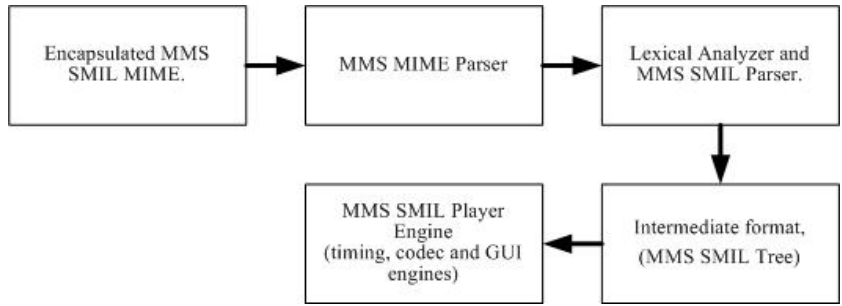

Figure 1: The Overall Architecture of MMS SMIL Player

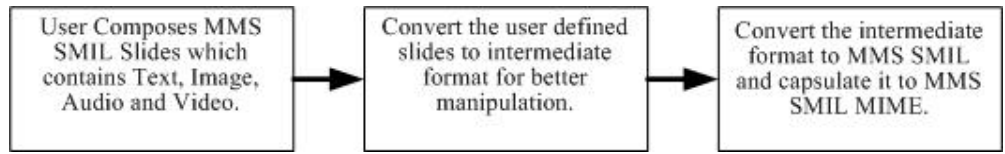

Figure 2: The Overall Architecture of MMS SMIL Composer

## 2. MMS Player and Composer Overall Architecture

The MMS message is an encapsulated MIME format message that means the SMIL and all the media are encapsulated into a file using MIME format. So this MIME file should be first parsed by a MIME parser to its components (i.e. texts, images, audios and videos). Then the SMIL itself is processed by the lexical analyzer. Lexical analyzer tokenizes the input stream to a stream of tokens and the parser converts this output stream to the parser tree called SMIL Tree. The SMIL Tree is the entry for SMIL player engine. SMIL player engine should go through the SMIL Tree and presents the tree.

The figure 2 shows the overall architecture of the MMS SMIL composer. The user constructs MMS message by creating several slides. A slide is created by specifying its text, image, audio or video. The second block converts these slides to the SMIL tree for better manipulation. The third block converts the MMS SMIL tree to the MMS SMIL or capsulated MMS SMIL MIME file.

As declared the SMIL tree is the basic data structure of the system. Both the player and composer use this structure as their core data structure. This makes the system robust, flexible and efficient. For example when composing a message the player engine easily could be incorporated for the previewing each slide without further coding. The importance of choosing this structure is high lighted especially when the system resides on mobile platform that the power supply and the operating system resources are limited. In the next session this data structure is introduced.

## 3. MMS SMIL Intermediate Format

SMIL tree structure is based on the elements in the MMS message SMIL. SMIL like other markup languages uses the XML standard, but has its own elements. Each SMIL file consists of a head and a body. In head there are the layout and the regions in which the media are presented. Body contains pars which resemble slides. Each par then has its own media. After parsing the SMIL file and before the message could be played by the player there is a preprocessing step which is called the layout fitting step. This preprocessing is responsible for adapting the layout and region dimensions encountered in MMS message with the dimensions of the device on which the message is being played.

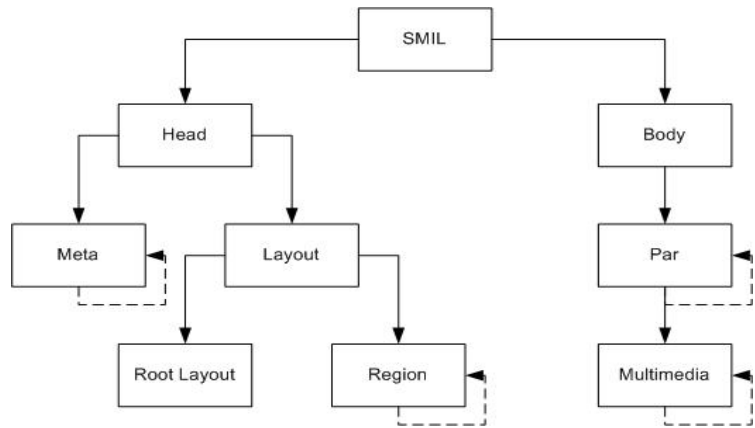

Figure 3: Intermediate Format Elements

This adaptation is performed according to the MMS conformance document version 2.0.0 that tells in what condition the device should do what. Figure 3 shows the MMS SMIL tree. After creation of the SMIL tree, this tree is passed to the player engine.

## 3. MMS SMIL Player Engine

SMIL supports both sequential and concurrent representations. Representing sequentially is simple but for supporting concurrent operation, each task has to be executed in the parallel form, so multitasking should be incorporated. Multitasking is done in OS layer, OS dependent APIs are used for handling multitasking. Note that for MMS SMIL there is at most one text, one image, one audio, one ref. and one video in parallel but this engine is capable of concurrently handling several multimedia objects in Z order basis. Figure 4 represents the general architecture for presentation engine.

Each par is a like a frame in a movie, but each frame has its own functionality. In a film each frame is presented with a constant rate, but here each par has its own start and duration time attribute. The player engine gets these attributes and starts presenting the par.

As said a par is an active frame, or it could be said that it is by itself a movie. But here each media in a par resembles a frame. Like a par each media also has its own start and duration time relative to the container par, but the par start time and duration dominate.

As it is shown in figure 4, for the each par and each multimedia, one timer is assigned. By the start message of the timer the multimedia is presented and with the end time message, the presentation will be stopped. This architecture is an event based one, so there is no extra load on CPU and is ideal especially for hand held devices. After presenting of the current par, the engine goes to the next par until the end of the tree.

Player engine also have several interfaces for supporting user interface functions, suppose the user interface may contain play, pause, stop, rewind and next buttons, so these functionalities are delivered to the user by some predefined interfaces.

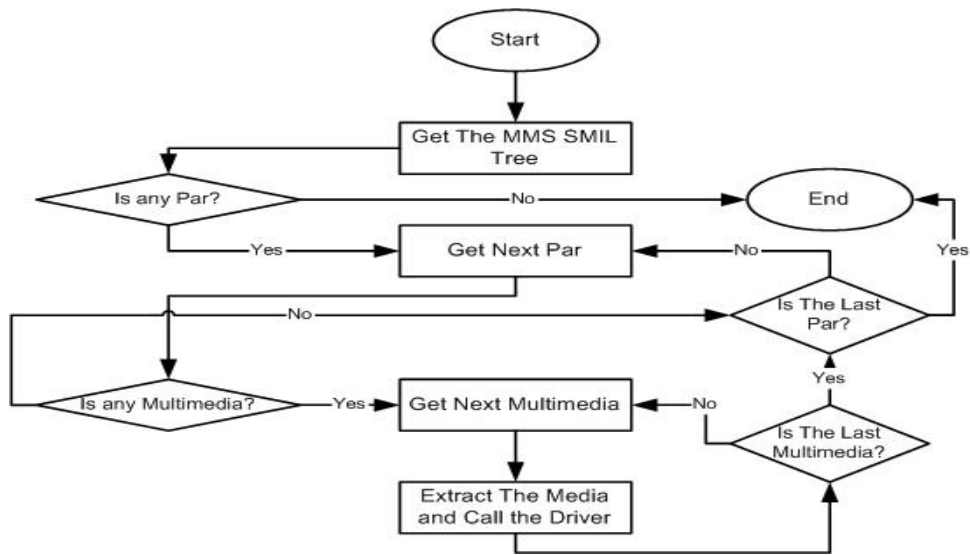

Figure 4: Timing Engine Flowchart

## 4. MMS SMIL Composer System Architecture

The main function of this unit is to make a flexible and easy to use tool that the MMS customer can make his/hers multimedia message. This unit should have flexible GUI.

MMS composer has a tight relation to the GUI and operating system APIs. For designing a multiplatform system the idea here is the separating the whole system into two subsystems, one for functionality and the other for the GUI interaction. When porting to the other platforms only the second should be adapted. Here the usage of the SMIL tree structure makes this architecture easy to get realized, as the first subsystem contains APIs for SMIL tree operations and the second interacts with GUI functions and events. After completion of the message composition the SMIL tree is converted to a MMS MIME message for transmission. This architecture was shown in figure 2. The use case diagram of this tool is shown in figure 5.

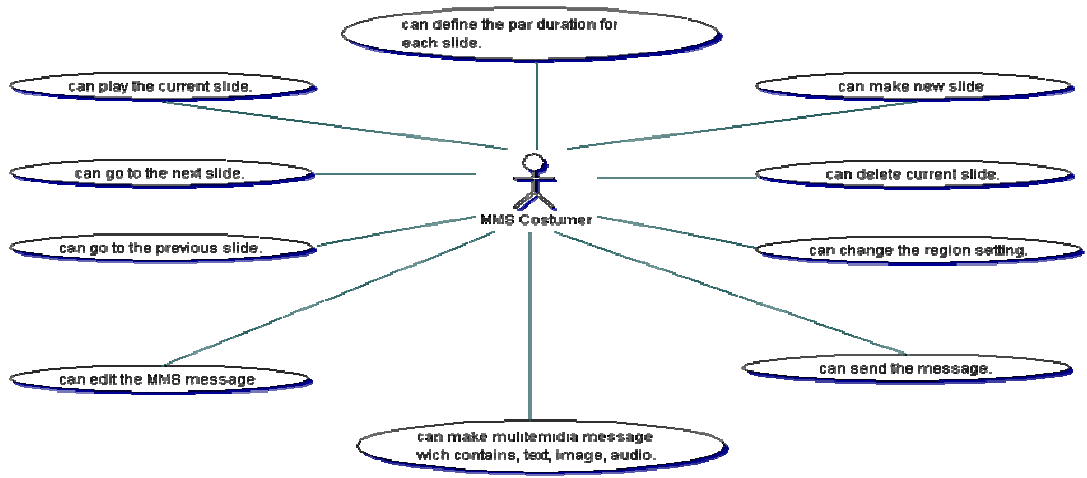

Figure 5: The Use Case Diagram for the MMS SMIL Composer

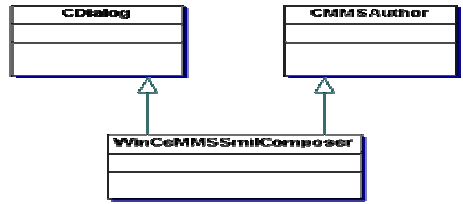

Figure 6: Proposed Methods for MMS SMIL Composer Architecture.

For combining the two systems there would be architectures as in figure 6. the WinCEMMSSMILComposer class inherits the CMMSAuthor which has SMIL tree APIs and CDialog which contains the GUI APIs and events.

## 5. MMS MIME Encapsulation and Decapsulation

MIME (**M**ulti purpose **I**nternet **M**ail **E**xtension) is a standard for transmission of any file by email on the internet. In MMS MIME is selected as standard for MMS message. MIME is a method for encapsulating different types of objects into a file. This is done by adding one file header and a body header for each object. These headers store some information about each object in the MIME file, also has a boundary string which plays as a separator between different objects. Figure 1 shows the capsulated MMS message. In this format the transport layer adds some headers to the message for example about day, time, user id, etc. The body contains the SMIL source file and the related multimedia object such as image, text, audio, and video [5,6].

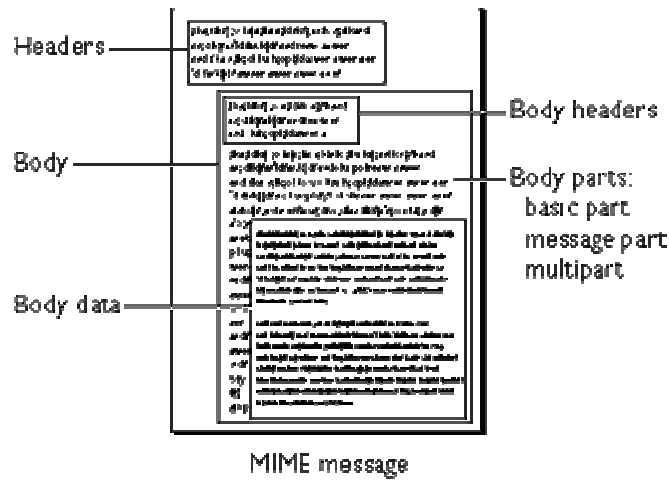

Figure 7: The MIME Message Format

According to the MIME grammar the MMS MIME needs a parser that can parse this grammar and decapsulates the MMS message. The functionality of the MMS MIME capsulator is easy, which adds some tags to the multimedia objects and then capsulate all together. Figure 7 shows the architecture of these engines.

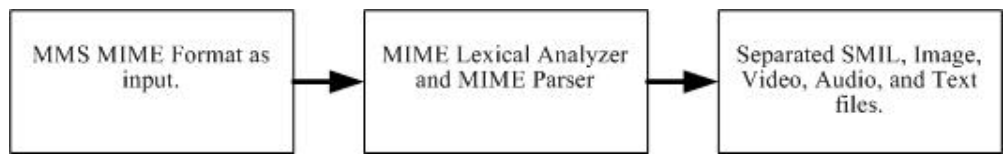

Figure 8: The Overall Architecture of the MIME Decapsulator

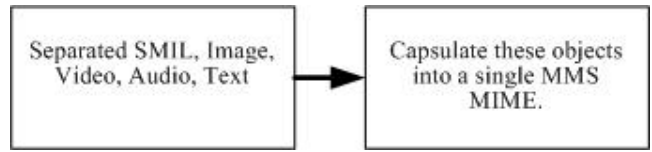

Figure 9: The Overall Architecture of the MIME Encapsulator

## 6. MMS Transport Layer

Like other communication systems MMS also have its own transport layer. Transport layer is responsible for handling the signaling needed to send a message to the Proxy-Relay Server and to receive a message from it. The Proxy-Relay Server is then responsible for delivering the sent message to the target. Transport layer has several commands and status attributes. Commands as SEND, DELETE, FORWARD and etc. say the proxy-relay server what to do with the message and status attributes tells the client what has done or happened to its message. For example for sending a message transport layer send a SEND command to the proxy-relay server and after the message was delivered successfully, transport layer notifies the user that the SEND command was successful. In figure 10 the transport layer in MMS system is shown. As it is shown there are 4 queues which solve the asynchrony
between the send and receive path.

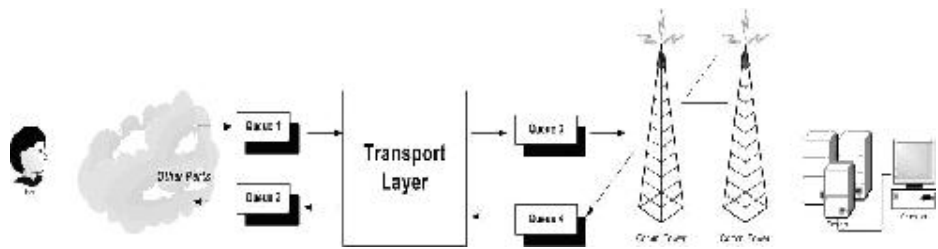

Figure 10: Transport layer in MMS system

## 7. MMS Proxy-Relay Server and MMS Client Emulator

For testing the overall MMS client system, a Proxy-Relay server and a MMS client Emulator has been developed. As the real system uses the WAP for sending and receiving the MMS message, these two systems were developed on ordinary LAN. The server emulator consists of transport layer, network layer, server manager and GUI.
Network layer acts as a system for sending and receiving MMS messages over the network. Here the transport layer is responsible for retrieving commands and statuses. After extraction commands and statuses, appropriate action is chosen by the server and the message is routed to the other clients and the necessary status is sent back to the sender client.

The server manager keeps track of the live clients with their IP and port; this enables to have more clients on the same machine with one IP address. If a client is off, the manager holds its message up to a dead time, waiting for the client to login to the network. There is no pooling system for tracking which client is off or on. Every time a client logs into the network, it registers itself and the manger sends the offline messages to it server GUI has specials panes for monitoring incoming and outgoing packets, also displays each packet command and status.

Client emulator consist of a complete MMS player and composer, transport layer, message manager pane, function keys and also monitoring panes as the server emulator. This client emulator fully emulates a handheld MMS enabled mobile phone.

The server address and port could be set and each incoming and outgoing packet could easily be traced. This system is ideal for testing such a detailed and sophisticated system. Figure 11 left shows the server and figure 12 shows the client emulator. Message manager is a graphical interface for sending, playing each message after being selected.

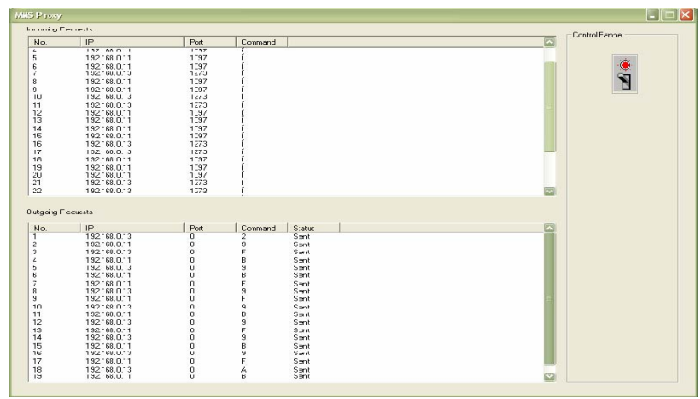

Figure 11: MMS Proxy-Relay Server Emulator

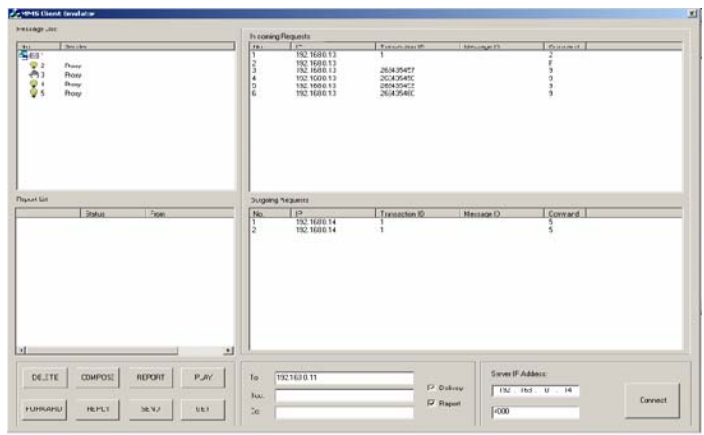

Figure 12: MMS Client Emulator

## 8. Conclusion

In this paper a detailed methodology for implementing a MMS client has been proposed. Each part is the result of so many research and trials. This system with this architecture was developed and passed so many tests. All the subsystems were considered on the emulators for testing.

The test procedure was a set up that consist of a server emulator running on a P4 PC with 256 MB Ram, and 7 clients running on the 7 PCs with the same specifications as server. The clients were restructured to send each 100 messages with different commands. There were two pocket PCs with WinCE platform as the 8'Th and 9'Th clients.

The server successfully was capable of handling average 500 requests per second. The response time differs in each command regarding to the actions needed to be performed. The number of request could be handled by the system mainly depends on the server resources specially servers' RAM that by doubling the RAM of the server the average request per second that could be handled also increased about 1.7 times when it was 256 MB.

This system was developed in Eimaa Communication Technologies Inc. as a solution for next generation of Iran's mobile system.